\documentclass[a4paper,fleqn]{cas-dc}
\usepackage{graphicx}	 % Including figure files
\usepackage{amsmath}	 % Advanced maths commands
\usepackage{amssymb}	 % Extra maths symbols
\usepackage{mathtools}   % multi-lines for equation
\usepackage[dvipsnames]{xcolor}      % comments color
\usepackage[authoryear]{natbib}
\usepackage{xurl}
\usepackage{hyperref}
\usepackage{ulem}

\usepackage{siunitx}
\usepackage{orcidlink}
\hypersetup{
    colorlinks=true,
    linkcolor=blue,
    filecolor=magenta,      
    urlcolor=blue,
    citecolor=blue,
    }

\newcommand{\Ms}{\mathrm{M_\odot}}

\begin{document}
\let\WriteBookmarks\relax
\def\floatpagepagefraction{1}
\def\textpagefraction{.001}

\shorttitle{CDM and SIDM Interpretations of Cloud-9 }

\shortauthors{Morgan Ohana et~al.}

\title [mode = title]{Cold Dark Matter and Self-Interacting Dark Matter Interpretations of Cloud-9}

\author[ceci]{Morgan Ohana}[
orcid=0009-0005-1165-504X
]
\cormark[1]

\ead{morgan.ohana@email.ucr.edu}

\credit{Investigation, Validation, Writing – review \& editing}

\author[ceci]{Xingyu Zhang}[
orcid=0009-0009-2791-1684
]

\ead{xingyu.zhang@ucr.edu}

\credit{Investigation, Validation, Writing – review \& editing}

\author[ceci]{Hai-Bo Yu}[
orcid=0000-0002-8421-8597
]

\ead{haiboyu@ucr.edu}
\credit{Conceptualization, Resources, Supervision, Validation, Writing – review \& editing}

\affiliation[ceci]{organization={Center for Experimental Cosmology \& Instrumentation, Department of Physics and Astronomy, University of California},
city={Riverside},
postcode={92521},
state={California},
country={USA}}

% Corresponding author text
\cortext[1]{Corresponding author}

\begin{abstract}
Recently, the Five-hundred-meter Aperture Spherical Telescope discovered a gas-rich hydrogen cloud near M94 in the $21\,{\rm cm}$ band. Lacking an optical counterpart, this object, dubbed Cloud-9, has been identified as a compelling Reionization Limited \textsc{Hi} Cloud (RELHIC). RELHICs provide exceptionally clean laboratories for probing dark matter, free from the baryonic complexities associated with star formation and feedback. We show that the observed hydrogen column density profile of Cloud-9 is consistent with a gas cloud embedded in either a cuspy halo predicted by the standard cold dark matter (CDM) model or a cored halo produced by self-interacting dark matter (SIDM). In both cases, the halo must have an unusually diffuse central density. The best-fitting CDM halo lies around $7\sigma$ below the cosmological concentration--mass relation, whereas SIDM core-forming halos reduce the tension to only around $3\sigma$. We further identify Cloud-9 analogs in the Concerto suite of cosmological zoom-in simulations with velocity-dependent SIDM, demonstrating that RELHICs provide a promising new probe of dark matter self-interactions.

\end{abstract}


\maketitle

\section{Introduction}

In the standard cosmological model, cosmic structure forms hierarchically, giving rise to a dark matter halo mass function that spans many orders of magnitude and rises steeply toward low masses \citep{Jenkins2001}. However, most low-mass halos remain invisible because there exists a critical mass below which a halo can no longer retain its gas and form a luminous galaxy. Simulations suggest this threshold is around $10^8\textup{--}10^9\,\Ms$ \citep{Nadler:2025mnz,benitez-llambay_2017}. Of particular interest are halos near this threshold that fail to form stars but retain a nearly spherical, hydrostatically supported hydrogen cloud in thermal equilibrium with the ionizing ultraviolet background. These objects, known as Reionization Limited \textsc{HI} Clouds (RELHICs), were first predicted by \citet{benitez-llambay_2017}. Recently, observations with the Five-hundred-meter Aperture Spherical Telescope (FAST) identified a promising RELHIC candidate, dubbed Cloud-9, in the vicinity of M94 \citep{Zhou2023}. Subsequent \textit{Hubble Space Telescope} observations detected no optical counterpart, further strengthening the case for Cloud-9 as the most compelling RELHIC candidate identified to date \citep{Anand2025}. \cite{Zhou:2026} further showed that the \textsc{HI} distribution exhibits a two-component structure, consisting of a compact core and an extended envelope. 

RELHICs provide a unique probe of low-mass dark matter halos because their gas cores are expected to be in thermal equilibrium with the ionizing UV background and hydrostatic equilibrium within the gravitational potential of their host halos. Therefore, measurements of the gas distribution in Cloud-9, such as those presented in \cite{Anand2025}, enable the underlying dark matter density profile to be inferred through the hydrostatic equilibrium equation. Furthermore, because RELHICs contain no stars, or at most only a negligible stellar component, they are largely free from the baryonic feedback processes that complicate the interpretation of dark matter distributions in ordinary galaxies. RELHICs therefore provide a clean testbed for testing models of dark matter.

In this work, we interpret the observations of Cloud-9 within the cold dark matter (CDM) and self-interacting dark matter (SIDM)~\citep{Tulin2018} frameworks. For CDM, we model the host halo with a Navarro--Frenk--White (NFW) density profile~\citep{Navarro:1996gj}, while for SIDM we employ the parametric halo model developed in \cite{Yang:2023jwn,ParametricModel}. For each set of halo parameters, we numerically solve the hydrostatic equilibrium equation to obtain the gas distribution, construct the corresponding \textsc{Hi} column density profile, and compare it with the Cloud-9 observations. We perform a Markov chain Monte Carlo (MCMC) analysis to reconstruct the posterior distributions of the halo parameters. In addition, we identify solutions that satisfy hydrostatic equilibrium but are dynamically unstable and exclude them from the inference. 

We demonstrate that, while the current Cloud-9 observations remain consistent with the cuspy density profile of an NFW halo, the inferred halo must be exceptionally diffuse, with a concentration lying around $7\sigma$ below the cosmological concentration--mass relation~\citep{Diemer2019}. Within the SIDM framework, our parameter exploration reveals a characteristic degeneracy between halo concentration and the stage of gravothermal evolution. Although halos at different evolutionary stages can reproduce the observed gas profile, those in the core-formation phase minimize the tension with the cosmological concentration--mass relation. For self-interacting cross sections per unit mass of $\sigma/m\gtrsim50\,{\rm cm^2/g}$, the preferred halos are only around $3\sigma$ below the cosmological median. Intriguingly, we further identify potential Cloud-9 analogs in the Concerto suite of cosmological zoom-in simulations with velocity-dependent SIDM~\citep{Nadler:2025jwh,Nadler:2023nrd}.

The rest of the paper is organized as follows. In Sec.~\ref{sec:method}, we describe our modeling framework for Cloud-9 in the CDM and SIDM frameworks. In Sec.~\ref{sec:results}, we present the main results, including the best-fit models, posterior distributions of the halo parameters, the correlation between halo concentration and the self-interacting cross section, and the hydrostatic stability criterion. In Sec.~\ref{sec:sims}, we identify dark matter halo analogs of Cloud-9 in cosmological zoom-in SIDM simulations.

\section{Method}
\label{sec:method}
We compute the gas density profile of Cloud-9 by solving the hydrostatic equilibrium equation,
\begin{equation}
\frac{dP}{dr} = - \frac{G M_{\rm enc}(r) \rho_{\rm gas}(r)}{r^2},
\label{eq:gas}
\end{equation}
where $P$ is the gas pressure, $\rho_{\rm gas}$ is the gas density, and $M_{\rm enc}$ is the enclosed mass within radius $r$, including contributions from both gas and dark matter. We model the gas as locally isothermal and adopt the ideal gas equation of state, $P=\rho_{\rm gas} (k T/\mu)$, where $T$ is the gas temperature, $k$ is the Boltzmann constant, and $\mu$ is the mean molecular weight. Using
$M_{\rm enc}(r)=4\pi\int_0^r dr' r'^2 \left(\rho_{\rm gas} + \rho_{\rm DM}\right)$, Eq.~\ref{eq:gas} can be rewritten as
\begin{equation}
\label{eq:diffeq}
\frac{d\rho_{\rm gas}}{dr} = - \frac{\left(\frac{4\pi G\mu}{kT}\right) \frac{\rho_{\rm gas}}{r^2}\int_0^r dr' r'^2 \left(\rho_{\rm gas} + \rho_{\rm DM}\right)}{1 + \frac{\partial \ln T}{\partial \ln \rho_{\rm gas}} - \frac{\partial \ln \mu}{\partial \ln \rho_{\rm gas}}},
\end{equation}
where $\rho_{\rm DM}$ is the dark matter density.

To render Eq.~\ref{eq:diffeq} solvable, we compute the temperature $T$ as a function of the gas density $\rho_{\rm gas}$ using the temperature--density relation from \citet{benitez-llambay_2017}, which was calibrated for RELHICs in cosmological simulations. We further apply the method in \citet{Rahmati2013} to determine the neutral hydrogen fraction, allowing the mean molecular weight $\mu$ to be computed self-consistently as a function of gas density and temperature. Therefore, all quantities on the right-hand side of Eq.~\ref{eq:diffeq} can be expressed in terms of the gas density for a given dark matter density profile, enabling the equation to be numerically integrated outward from the center for a given central gas density $\rho_c$. The resulting gas profile is then used to reconstruct the neutral hydrogen column density, which is compared with the observational data from \citet{benitez-llambay_examining_2024} (their Fig.~4). To accurately capture the gas distribution over the full extent of the halo, we integrate from $15\,{\rm pc}$ to the radius at which the dark matter density equals the critical density of the Universe.

For CDM, we model the dark matter halo with an NFW density profile~\citep{Navarro:1996gj},  
\begin{equation}
\label{eq:nfw}
\rho(r) = \frac{\rho_s}{ \frac{r}{r_s}\left(1 + \frac{r}{r_s}\right)^2}
\end{equation}
where $r_s$ and $\rho_s$ are scale radius and density, respectively. 

For SIDM, we use the parametric SIDM halo model~\citep{Yang:2023jwn,ParametricModel}, where the density profile is given by
\begin{equation}
\label{eq:rhoSIDM}
\rho(r) = \frac{\rho_s}{\frac{\left(r^4 + r_c^4\right)^{1/4}}{r_s}\left(1 + \frac{r}{r_s}\right)^2},
\end{equation}
where $r_c$ is the core radius. The three model parameters \(\rho_s, r_s\), and $r_c$ are functions of the dimensionless evolution timescale $\tau$ as 
\begin{equation}
\label{eq:parametric_evol}
\begin{aligned}
    \frac{\rho_s}{\rho_{s,0}} &= 2.033 + 0.7381 \tau + 7.264 \tau^5 - 12.73 \tau^7 + 9.915 \tau^9 \\
    &+ (1 - 2.033) \frac{\ln(\tau + 0.001)}{\ln(0.001)} \\
    \frac{r_s}{r_{s,0}} &= 0.7178 - 0.1026 \tau + 0.2474 \tau^2 - 0.4079 \tau^3 \\
    &+ (1 - 0.7178) \frac{\ln(\tau + 0.001)}{\ln(0.001)} \\
    \frac{r_c}{r_{s,0}} &= 2.555 \sqrt{\tau} - 3.632 \tau + 2.131 \tau^2 - 1.415 \tau^3 + 0.4683 \tau^4 \\
\end{aligned}
\end{equation}
where $\rho_{s,0}$ and $r_{s,0}$ are scale density and radius of the corresponding {\it initial} NFW profile, respectively. The evolution parameter $\tau=t/t_c$ is the ratio of the time elapsed $t$ to the collapse timescale $t_c$ calculated as~\citep{Balberg:2002ue,Essig:2018pzq}
\begin{equation}
    \label{eq:t_c}
    t_{c} = \frac{150}{0.75} \frac{1}{(\sigma /m) \rho_{s,0} r_{s,0}}\frac{1}{\sqrt{4\pi G \rho_{s,0}}},
\end{equation}
where $\sigma/m$ is the self-interacting cross section per unit mass. In the limit $\tau=0$, we have $\rho_s=\rho_{s,0}$, $r_s=r_{s,0}$, and $r_c=0$, the density profile in Eq.~\ref{eq:rhoSIDM} reduces to the NFW profile in Eq.~\ref{eq:nfw}. Equivalently, we also specify a halo in terms of its mass $M_{200}$ and concentration $c_{200}$ following the relations
\begin{equation}
\label{eq:rhosrs}
\rho_{s,0}=\frac{\frac{200}{3}c^3_{200}\rho_{\rm crit}}{\ln(c_{200} + 1)-\frac{c_{200}}{(c_{200}+1)}},~r_{s,0}=\left(\frac{3M_{200}}{800\pi c^3_{200}\rho_{\rm crit}}\right)^{1/3}, 
\end{equation}
where $\rho_{\rm crit}$ is the critical density of the Universe. 

We explore the model parameter space using the MCMC method with flat priors. The halo mass is restricted to $10^8\, \Ms\leq M_{200}\leq 5\times10^9\, \Ms$, corresponding to the expected mass range of RELHIC host halos~\citep{benitez-llambay_2017}, while the concentration is allowed to vary over $0\leq c_{200}\leq20$. The collapse parameter is sampled over $0\leq\tau\leq1$, where the upper limit corresponds to a halo in the deep core-collapse phase. The central gas density $\rho_c$ was allowed to vary within $8\times 10^4 \, \Ms /\textrm{kpc}^{3} \leq \rho_c \leq 1.6 \times 10^5 \, \Ms /\textrm{kpc}^{3} $. The exact choices of the prior bounds on $c_{200}$ and $\rho_c$ have a negligible impact on the results, as the observational constraints confine the posterior well within these limits.

We relate a given set of halo parameters $M_{200}$, $c_{200}$, and $\tau$ to the self- cross section $\sigma/m$ via Eq.~\ref{eq:t_c}, which can be rearranged as
\begin{equation}
    \label{eq:sigma_relation}
    t (\sigma/m) = \frac{150}{0.75} \frac{\tau}{\rho_{s,0} r_{s,0}}\frac{1}{\sqrt{4\pi G \rho_{s,0}}}
\end{equation}
This expression reveals a degeneracy between the halo age $t$ and the cross section $\sigma/m$. Throughout this work, we adopt $t=t_{\rm age}=10\,{\rm Gyr}$ as a fiducial halo age. Since the relation is linear, the inferred cross section can be readily rescaled for a different age assumption. In addition, we perform MCMC analyses at fixed cross section to further investigate the correlation between $c_{200}$ and $\sigma/m$ (Fig.~\ref{fig:sigma_vs_deviation}).

We perform our MCMC exploration using the affine-invariant stretch-move algorithm~\citep{Goodman:2010dyf}. We employ $512$ walkers, each initialized at a random point within the prior bounds. After a burn-in phase of $1000$ steps, sufficient to erase memory of the initial conditions, each walker is evolved for an additional $100,000$ steps, yielding a total of $5.12\times10^7$ posterior samples. For every analysis presented in this work, we verify convergence by requiring the split Gelman--Rubin statistic to be below $1.01$.

It is important to note that the hydrostatic equilibrium equation, Eq.~\ref{eq:gas}, guarantees equilibrium solutions for the gas density profile, but not all such solutions are dynamically stable. According to the standard turning-point criterion, stability requires the gas mass to increase monotonically with the central density, i.e., $dM_{\rm gas}/d\rho_c>0$~\citep{Friedman1988}. We impose this criterion and exclude unstable solutions in post-processing. In practice, it provides an effective lower bound on the halo mass $M_{200}$, as solutions with $M_{200}\lesssim2.5\times10^9\,\Ms$ are generally unstable. We discuss this stability criterion in more detail in Sec.~\ref{sec:stable}.

\section{Results}

\label{sec:results}

\subsection{Best Fits and Posterior Distributions}

\begin{figure*}
    \centering
    \includegraphics[width=0.5\linewidth]{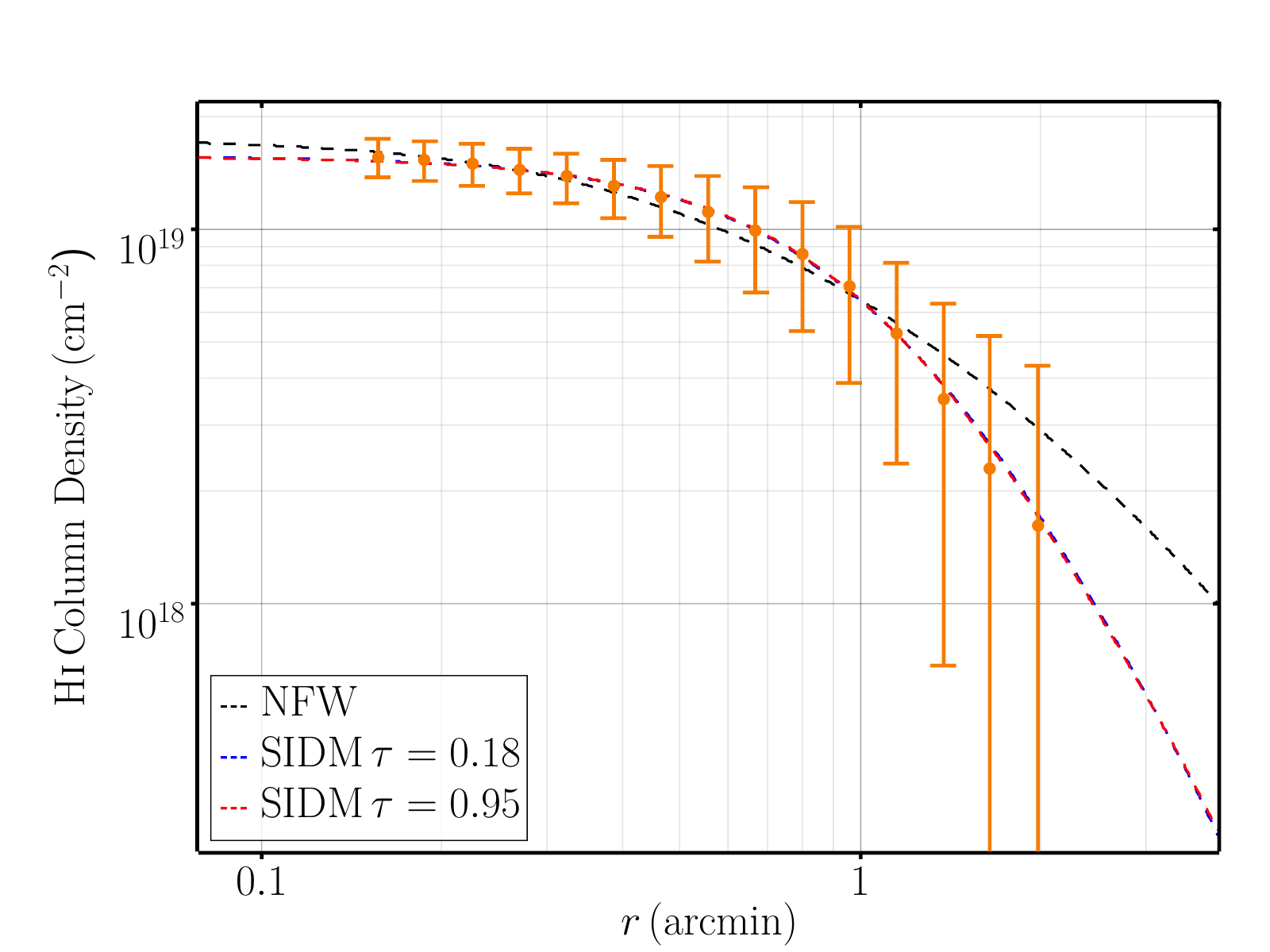}\includegraphics[width=0.5\linewidth]{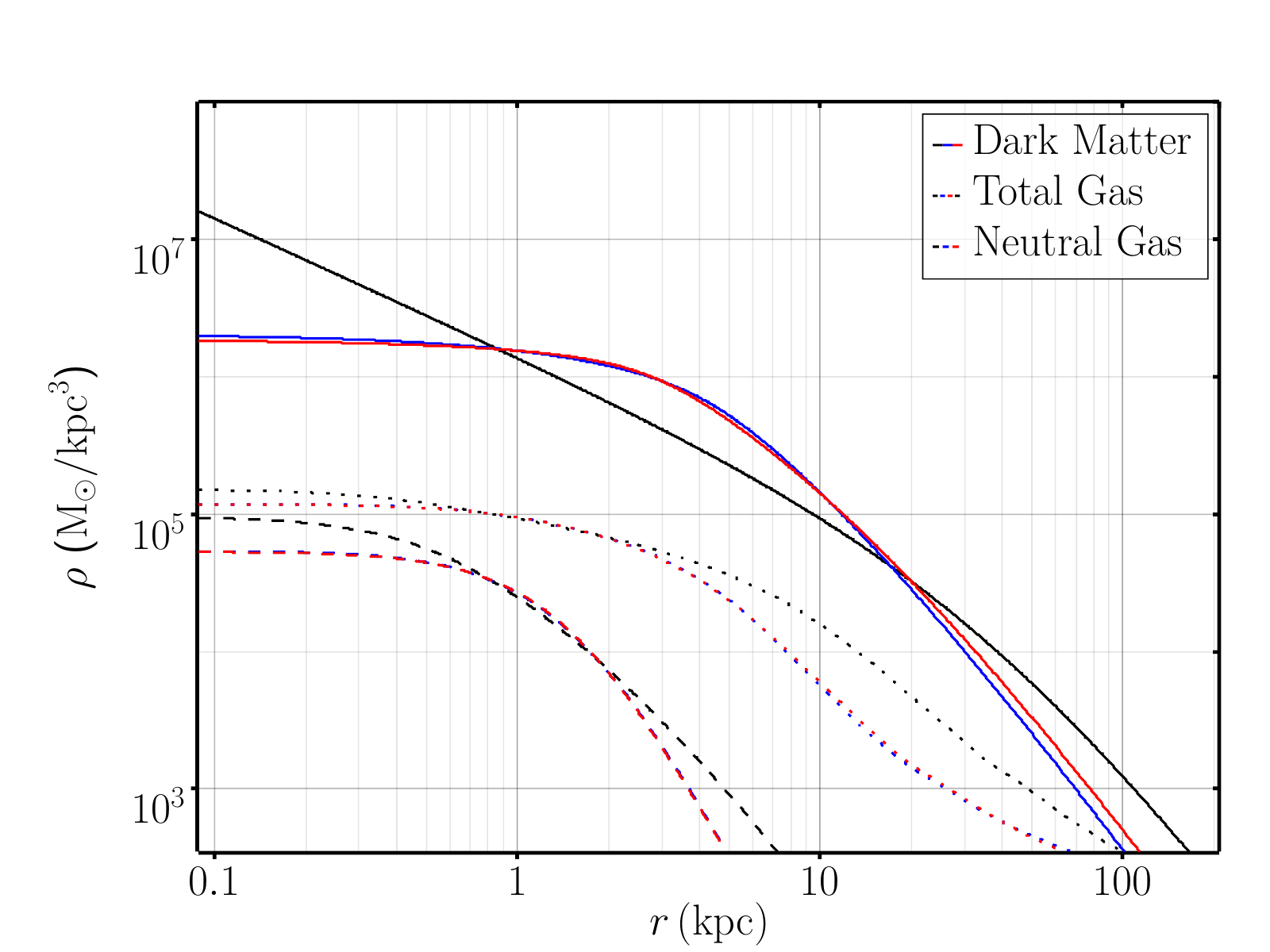}
    \caption{Three representative fits at different stages of gravothermal evolution that demonstrate that good fits can be obtained throughout the SIDM degeneracy. The left panel shows the neutral gas column density profiles compared with the observational data for an NFW halo (black), a core-forming SIDM halo (blue), and a core-collapsed SIDM halo (red). The right panel shows the corresponding three-dimensional density profiles of dark matter (solid), total gas (dotted), and neutral gas (dashed) for the three cases. The data points in the left panel are taken from Fig.~4 of \citet{benitez-llambay_examining_2024}.}
    \label{fig:fits}
\end{figure*}

\begin{figure*}
    \centering
    \includegraphics[width=\linewidth]{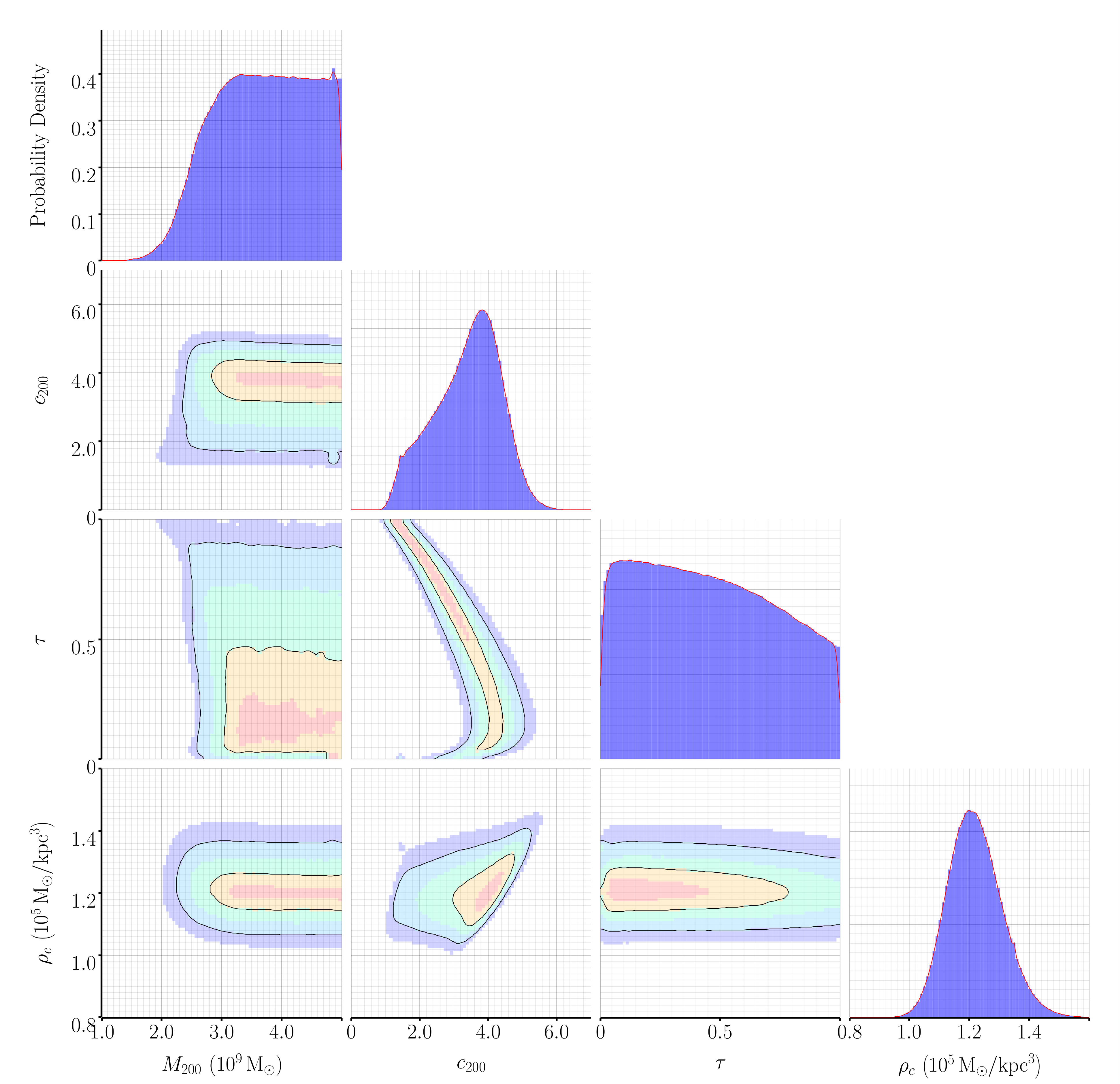}
    \caption{Corner plot showing the posterior distribution from the MCMC parameter space exploration. In the contour plots, each color transition corresponds to a change of $0.5\sigma$ in enclosed probability, while the solid black contours indicate the $1\sigma$ and $2\sigma$ credible regions.}
    \label{fig:cornerplot}
\end{figure*}

Our MCMC parameter exploration successfully identifies many good fits to Cloud-9, three of which are shown in Fig.~\ref{fig:fits}. The left panel compares the observed \textsc{Hi} column density profile from~\citet{benitez-llambay_examining_2024} (orange points) with the best-fitting NFW halo (black) and two SIDM halos with $\tau=0.18$ (blue) and $0.95$ (red). The two SIDM halos, despite representing markedly different stages of gravothermal evolution, produce nearly identical \textsc{Hi} column density profiles and provide excellent fits to the data, illustrating a characteristic degeneracy. The NFW halo also yields a good fit within the observational uncertainties. However, as we discuss below, although the NFW and SIDM models reproduce the Cloud-9 observations comparably well, the NFW interpretation requires an exceptionally low halo concentration.

The right panel in Fig.~\ref{fig:fits} shows the corresponding 3D density profiles of dark matter (solid), total gas (dotted), and neutral gas (dashed). The two SIDM halos exhibit nearly identical 3D density profiles, both featuring cores of size $\sim2\,{\rm kpc}$. In contrast, the NFW halo retains a central density cusp. The total gas distribution is more extended than the neutral gas distribution, while both remain more than two orders of magnitude below the dark matter density throughout the halo. Thus, Cloud-9 is strongly dark matter dominated, and the gravitational effects of baryons are negligible. This justifies modeling the CDM halo with the NFW profile and the SIDM halo with the parametric dark matter profile without explicitly accounting for baryonic effects~\citep{Kaplinghat:2013xca,Yang:2024tba,Jia:2026ocr,Kong:2026piq}.

To further investigate the degeneracy of the SIDM fits, we reconstruct the corresponding halo parameters from the MCMC chain. For the halo with $\tau=0.18$, we obtain $M_{200}=4.7\times10^9\,\Ms$ and $c_{200}=4.0$, corresponding to a $3.2\sigma$ deviation below the cosmological median concentration, assuming a scatter of $0.16$ dex~\citep{Diemer2019}. Using Eq.~\ref{eq:sigma_relation}, we infer a self-interacting cross section of $\sigma/m=483\,{\rm cm^2/g}$. Since an SIDM halo reaches its maximum core expansion at $\tau\approx0.146$, where its central density is minimized~\citep{Roberts:2024uyw}, this halo is close to the maximum core-expansion stage. For the second SIDM halo with $\tau=0.95$, we obtain $M_{200}=3.4\times10^9\,\Ms$ and $c_{200}=1.5$, corresponding to a $6.0\sigma$ deviation below the median. Although this deeply core-collapsed halo has a density profile comparable to that of the halo near maximum core expansion, it requires an extremely large self-interacting cross section of $\sigma/m=2.1\times10^4\,{\rm cm^2/g}$. This follows naturally from the scaling relation $t_c\propto(\sigma/m)^{-1}c^{-7/3}_{200}M^{-1/3}_{200}$~\citep{Essig:2018pzq,Zeng:2021ldo,Nadler:2023nrd}. Such a degeneracy is expected because the gravothermal evolution of SIDM halos follows a universal evolutionary track~\citep{Outmezguine:2022bhq,Zhong:2023yzk}. On the other hand, this result also implies that if the  cross section is extremely large, $\sigma/m\gtrsim10^4\,{\rm cm^2/g}$, reproducing Cloud-9 requires an exceptionally low-concentration halo. Such a scenario is therefore also disfavored, much like the CDM interpretation.

Fig.~\ref{fig:cornerplot} shows the corner plot that demonstrates overall constraints on the model parameters. The halo mass posterior peaks at $M_{200}\approx3.3\times10^{9}\,\Ms$, but exhibits a long, nearly flat tail toward higher masses, indicating that $M_{200}$ is only weakly constrained. As indicated in Eqs.~\ref{eq:rhoSIDM} and \ref{eq:parametric_evol}, the density profile is simply given by $\rho(r)\approx\rho_s r_s/r_c\propto\rho_{s,0}$ for $r\ll r_c\ll r_s$, corresponding to a constant-density core. Since $\rho_s$ depends explicitly on $c_{200}$, but not $M_{200}$ (Eq.~\ref{eq:rhosrs}), the fit loses sensitivity to $M_{200}$ once the core becomes large enough to encompass all the data points. For this reason, we impose a sharp cutoff at $M_{200}=5.0\times10^9\,\Ms$, motivated by \cite{benitez-llambay_2017}, since larger halos will likely form stars and thus are not good candidates for halos hosting RELHICs.

The concentration posterior peaks around $c_{200}\approx3.9$ and is nearly independent of the halo mass. For halos with $M_{200}\approx4\times10^9\,\Ms$, the expected median concentration is $c_{200}\approx13$ with a scatter of $0.16$ dex~\citep{Diemer2019}. Thus, the Cloud-9 halo favors a low concentration, corresponding to a $3.3\sigma$ deviation below the cosmological median. This is also reflected by the relatively extended tail of the $c_{200}$ posterior toward lower concentrations. We will come back to this point later.

The $c_{200}\textup{--}\tau$ posterior exhibits a thin, banana-shaped contour, reflecting the nonmonotonic evolution of the central density during gravothermal evolution. At both low and high $\tau$, corresponding to the initial NFW-like and deeply core-collapse phases, respectively, the halo needs a relatively low concentration to maintain a profile similar to core-forming halos of moderate $\tau$ and higher concentration. By contrast, at intermediate $\tau$, when the halo develops a dilute core, higher concentrations are permitted. The contour is broadest near $\tau\approx0.15$ and narrows rapidly toward both smaller and larger $\tau$, indicating that Cloud-9 mildly favors a diffuse cored halo, which is consistent with Fig.~\ref{fig:fits}. Although acceptable fits exist for both NFW and deeply core-collapsed halos, they require more finely tuned halo concentrations, as reflected by the narrowing of the posterior.

This conclusion is further supported by the marginalized $\tau$ distribution, which peaks near $\tau\approx0.1$, close to $\tau=0.146$, where an SIDM halo reaches its lowest core density. Note that the peak at $\tau\approx0.1$ does not imply that the best-fitting models occur exclusively at this stage. Indeed, Fig.~\ref{fig:fits} shows that comparably good fits are obtained over a broad range of $\tau$. Rather, it indicates that acceptable fits occupy a larger volume of parameter space near the maximum-core stage than in the NFW or deeply core-collapsed regimes, where the parameters require greater fine tuning. Lastly, we examine the central gas density and find that it is well constrained to $\rho_c\approx1.2\times10^5\, \Ms/{\rm kpc^3}$, while remaining largely insensitive to the halo mass $M_{200}$, as expected. Consistent with the previous results, the $\rho_c\textup{--}c_{200}$ and $\rho_c\textup{--}\tau$ posteriors favor a low halo concentration and an intermediate value of $\tau$.

\subsection{Cross Section and Halo Concentration}

\begin{figure}
    \centering
    \includegraphics[width=\linewidth]{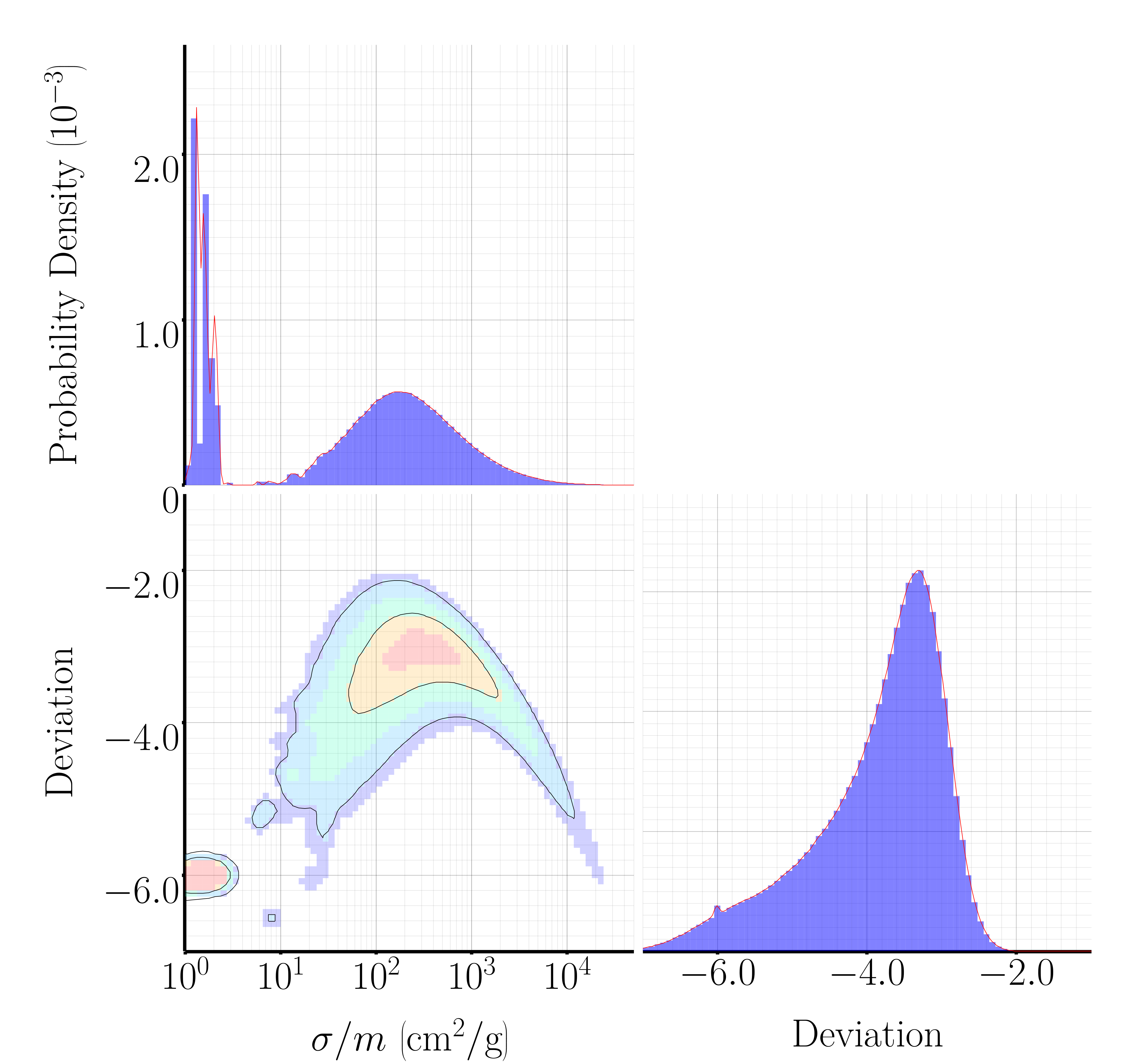}
    \caption{For each MCMC sample, we compute the corresponding SIDM cross section and deviation of concentration from the cosmological concentration--mass relation. The histograms of these quantities and their correlations are shown in the corner plot above. In the contour plots, each color transition corresponds to a change of $0.5\sigma$ in enclosed probability, while the solid black contours indicate the $1\sigma$ and $2\sigma$ regions.}
    \label{fig:minicorner}
\end{figure}

We further convert each sample from the MCMC chain into a cross section using Eq. \ref{eq:sigma_relation} by fixing the age to be $t=10\,{\rm Gyr}$, thereby reconstructing the posterior distribution of the SIDM cross section. In addition, we use the concentration--mass relation and its scatter from \citet{Diemer2019} to convert each sample into a deviation from the cosmological median concentration.

As shown in the top panel of Fig.~\ref{fig:minicorner}, the cross section posterior is predominantly a strongly right-skewed approximately log-normal distribution, with a peak at $\sigma/m$ around $200\,{\rm cm^2/g}$. This distribution disfavors both very small and very large cross sections while assigning substantial probability over a broad range of intermediate values, indicating that the data favor a diffuse cored halo. The bottom left panel of Fig.~\ref{fig:minicorner} shows the correlation between the cross section and the halo concentration, quantified by the deviation from the cosmological median. The posterior forms a clear arc, with the magnitude of concentration deviation minimized at approximately $3\sigma$ below the median for $\sigma/m$ around $200\,{\rm cm^2/g}$. Moving toward either smaller or larger cross sections rapidly requires increasingly low halo concentrations. The bottom right panel further shows that the concentration posterior peaks around $3.3\sigma$ below the cosmological median and exhibits a long tail toward even lower concentrations, again indicating that the data favor a substantially diluted halo. 

There is also a spike at very low cross section corresponding to a population of nearly CDM halos with high mass and low concentration. This population appears in the correlation plot as a peak localized near $6\sigma$ below the cosmological median, reinforcing the notion that cross sections in either extreme require extremely diffuse halos. These can also be seen in Fig.~\ref{fig:cornerplot} as the small local probability density spike in the bottom right of both the $M_{200}\textup{--}c_{200}$ and $M_{200}\textup{--}\tau$ correlation plots. Despite appearing like a disconnected peak in the posterior, this population has density profiles essentially the same as those already shown in Fig.~\ref{fig:fits}.

\begin{figure}
    \centering
    \includegraphics[width=\linewidth]{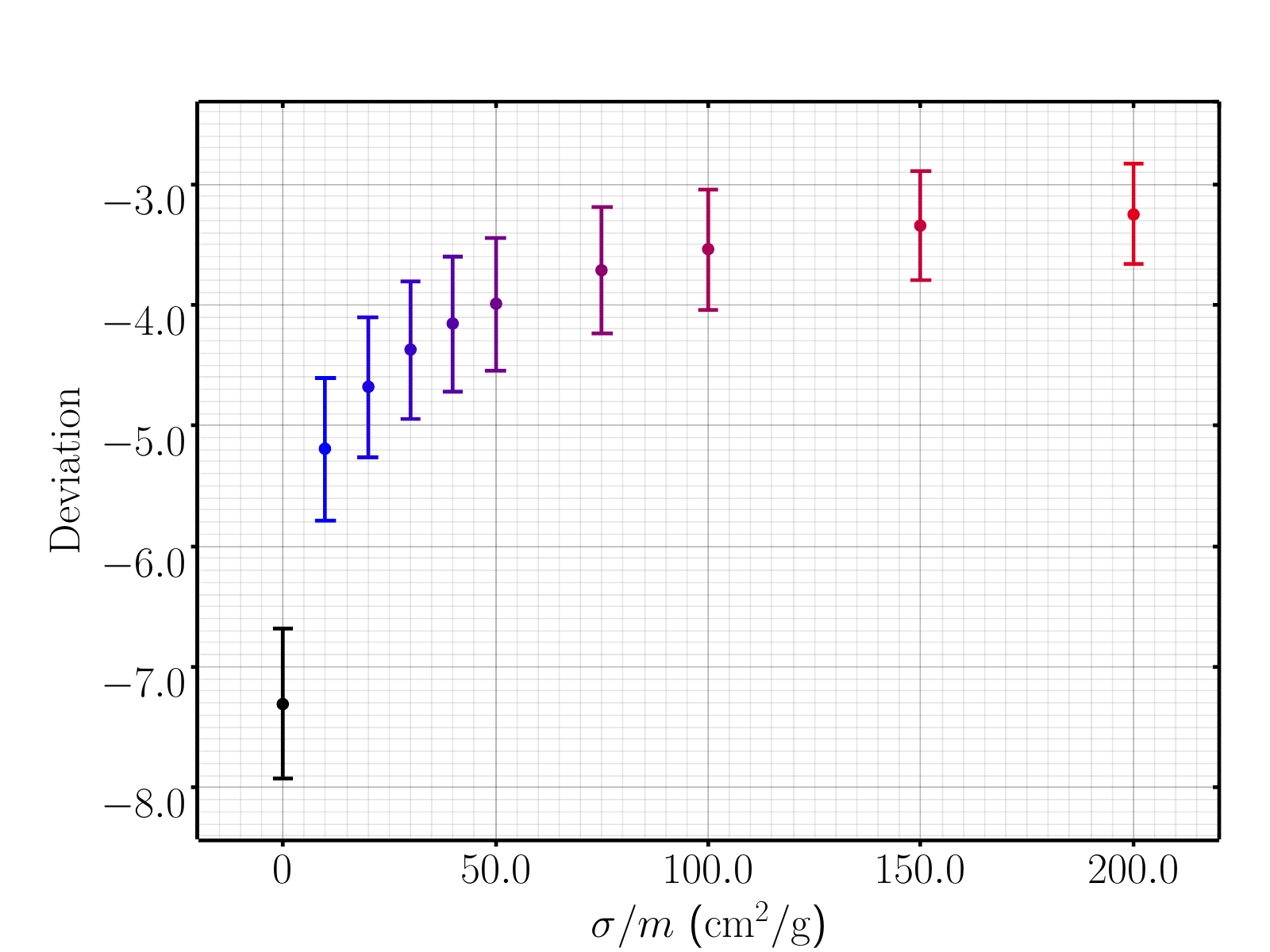}
    \caption{The deviation from the cosmological concentration--mass relation favored by each MCMC run as a function of the fixed self-interacting cross section. The error bars indicate the standard deviation of the posterior distribution of the concentration deviation for each MCMC run.}
    \label{fig:sigma_vs_deviation}
\end{figure}

Fig.~\ref{fig:minicorner} has a limited resolution at small cross sections, because such models are in general only weakly favored by the original MCMC exploration. To better probe this regime, we perform an additional set of MCMC analyses with the cross section fixed at a series of values and reconstruct the posterior distribution of the halo concentration deviation from the cosmological median at each cross section, as shown in Fig.~\ref{fig:sigma_vs_deviation}. We see that the required concentration deviation decreases rapidly with increasing cross section, until it levels off at around $3\sigma$ below the cosmological median for $\sigma/m\gtrsim50\textup{--}100\,{\rm cm^2/g}$. In the CDM limit, where $\sigma/m\rightarrow0$, the best-fitting halo requires a concentration $7.3\sigma$ below the cosmological median. Even allowing for the uncertainty in the fit, the required deviation remains larger than $6.7\sigma$. Thus, while CDM requires an exceptionally low-concentration halo to explain Cloud-9, SIDM naturally alleviates this tension through core formation.

\subsection{Hydrostatic Stability}
\label{sec:stable}

As discussed in Sec.~\ref{sec:method}, not all the equilibrium solutions found in our MCMC search are dynamically stable. In post-processing, we have imposed the criterion for stable gas configurations based on the turning-point method so the criterion for stability is $dM_{\rm gas}/d\rho_c > 0$~\citep{Friedman1988}, which provides an effective lower bound on the halo mass $M_{200}\gtrsim2.5\times10^9\,\Ms$. In this section, we provide more details and elaborate the physical reasons for the emergence of unstable solutions. 

For any hydrostatic system with a sufficiently deep confining potential, provided here by the gravitational field of the dark matter halo, both the total gas mass and the spatial extent of the gas cloud are finite. Thus, $M_{\rm gas}(\rho_{c})$ is a well-defined function. Through Eq.~\ref{eq:diffeq}, $M_{\rm gas}(\rho_{c})$ also depends on the underlying dark matter density profile. To illustrate this dependence, we consider two representative halos with parameters $(M_{200},c_{200},\tau)=(4.3\times10^9\,\Ms,4.1,0.18)$ and $(1.3\times10^9\,\Ms,2.2,0.87)$, corresponding to a stable and an unstable solution, respectively. For each halo, we solve Eq.~\ref{eq:diffeq} over a range of central gas densities to determine the corresponding total gas mass.

\begin{figure}
    \centering
    \includegraphics[width=\linewidth]{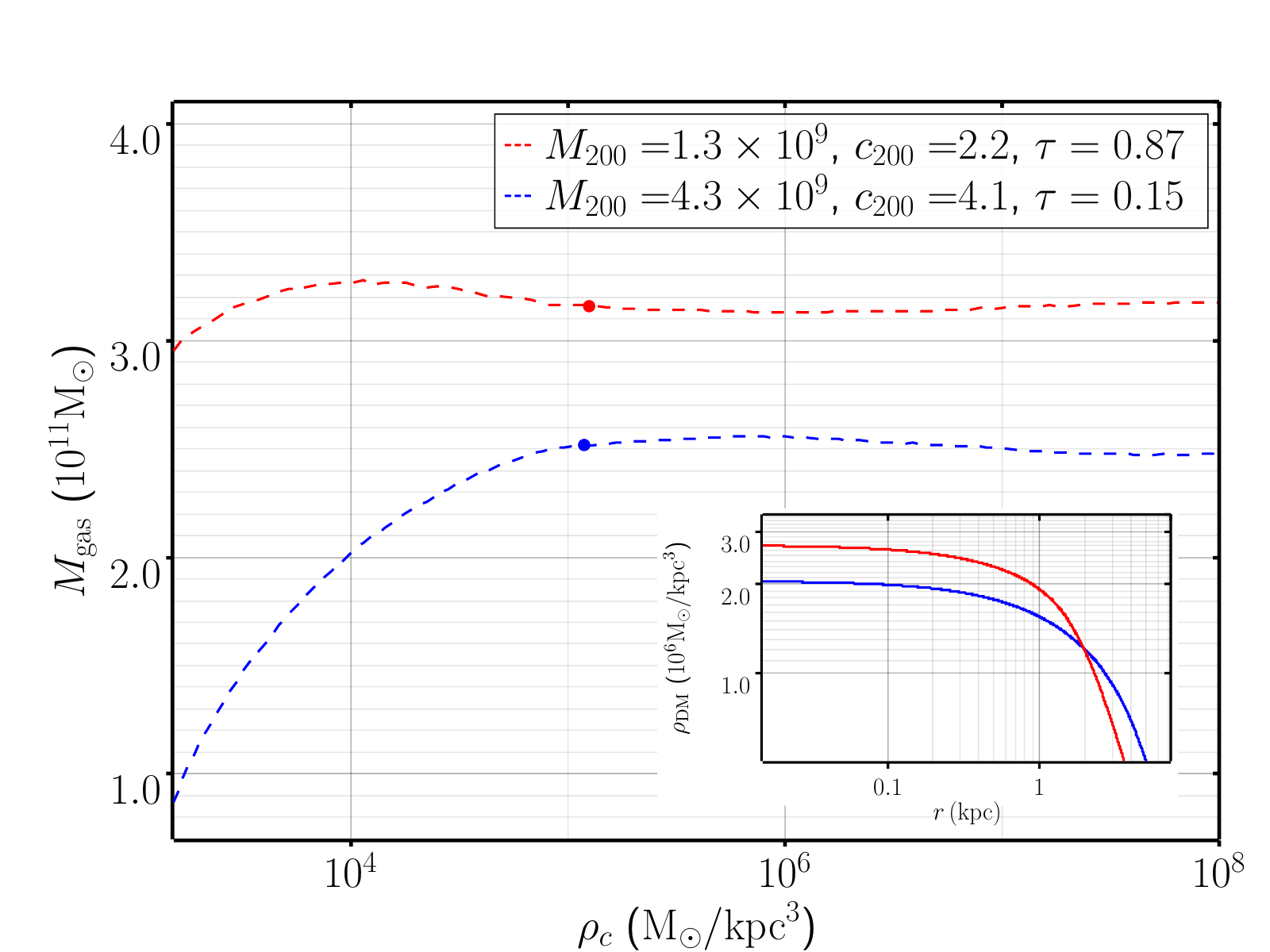}
    \caption{The total gas mass as a function of the central gas density for two similar fits for halos with $M_{200}=4.3\times10^9\,\Ms$ (blue) and $1.3\times10^9\,\Ms$ (red), representing two sides of the critical region around $M_{200} = 2.5 \times 10^9\,\Ms$. The dot on each curve is where the Cloud-9 fit lies. For the blue curve, the solution lies on the stable branch $dM_{\rm gas}/d\rho_c>0$. However, for the red curve a maximum has appeared and the solution lies on the unstable branch where $dM_{\rm gas}/d\rho_c<0$.}
    \label{fig:instability}
\end{figure}

The main panel of Fig.~\ref{fig:instability} illustrates the gas mass as a function of the central density for the halos with $M_{200}=4.3\times10^9\,\Ms$ (blue) and $1.3\times10^9\,\Ms$ (red), together with corresponding solutions for Cloud-9 (dots). For the case denoted by the blue curve, $M_{\rm gas}$ increases monotonically with $\rho_c$ and then saturates. Apparently, for the corresponding solution to Cloud-9 denoted by the blue dot, the stable condition holds, i.e., $dM_{\rm gas}/d\rho_c>0$. On the other hand, for the case denoted by the red curve, $M_{\rm gas}$ first increases monotonically with $\rho_c$, reaches maximum, and then decreases. For the corresponding solution, $dM_{\rm gas}/d\rho_c<0$, and hence it is unstable.  

The stability criterion can also be understood from the behavior of nearby nonequilibrium configurations. At fixed central density, solutions below the equilibrium curve are under-massive, so gravity is too weak to balance the gas pressure and the cloud expands toward lower central density. Conversely, solutions above the curve are over-massive and contract toward higher central density. Thus, when the equilibrium curve has a positive slope, perturbations on either side drive the system back toward equilibrium, implying dynamical stability. When the slope is negative, the perturbations instead grow, rendering the equilibrium unstable. As illustrated in Fig.~\ref{fig:instability}, an expanding unstable configuration eventually reaches the stable branch of the equilibrium curve, where the evolution halts. In contrast, a contracting configuration undergoes runaway collapse until the assumptions underlying the RELHIC equilibrium model break down. The ultimate outcome may involve star formation or other compact objects, which we leave for future investigation.

The inset panel of Fig.~\ref{fig:instability} shows the dark matter density profiles for the two cases illustrated in the main panel. It is clear that the one with the smaller halo mass $M_{200}=1.3\times10^9\,\Ms$ have a high inner density.  Recall that the inner density is controlled by $\rho_{s,0}$, which is sensitive to $c_{200}$, but not $M_{200}$. Halo mass is primarily determined by physical extent, so low $M_{200}$ does not mean low density, but rather more compact and thus overall less massive halos. A natural interpretation of this is then that a gas profile consistent with the data from Cloud-9 cannot find stable support in the steep compact potentials of low-mass halos. Filtering through all fits from our MCMC scan we find that $50\%$ of them are unstable with $dM_{\rm gas}/d\rho_c<0$. In particular, there is a sharp transition around $M_{200}=2.5 \times 10^9\,\Ms$, below which most of the fits are unstable. This is evident from the histogram of the halo mass in Fig.~\ref{fig:cornerplot}, where we remove all the unstable solutions.  

\section{Realizations in cosmological simulations}
\label{sec:sims}
Using the dark matter density profiles for Cloud-9 inferred from our MCMC analysis, we search for analogous halos in the Concerto suite of cosmological zoom-in SIDM simulations~\citep{Nadler:2023nrd}, whose CDM counterparts are provided by the Symphony suite~\citep{Nadler:2022dvo}. The Concerto suite includes host halos spanning masses from $10^{11}\,\Ms$ to $10^{14}\,\Ms$ and adopts three velocity-dependent SIDM models. In this work, we analyze the Halo004 simulation, whose host halo has a mass of $\sim10^{12}\,\Ms$. This simulation employs the GroupSIDM-147 model~\citep{Nadler:2023nrd} which assumes a Yukawa-like dark matter interaction~\citep{Feng:2009hw,Tulin:2013teo} with a cross-section normalization of $\sigma_0/m=147.1\,{\rm cm^2/g}$ and a characteristic velocity of $w=120\,{\rm km/s}$. It has been shown to produce a wide diversity of dark matter density profiles through gravothermal expansion and collapse, potentially explaining the observed diversity of galactic dark matter distributions~\citep{Nadler:2023nrd,ManceraPina:2024ybj,Yu:2025tmp,Kong:2025sqx,Zhang:2026gur,Schmidt:2026hmn}. The zoom-in region has a particle mass of $5.0\times10^4\,\Ms$ and a gravitational softening length of $\epsilon=114\,{\rm pc}$, providing sufficient resolution to identify Cloud-9 analogs.  

\begin{figure*}
    \centering
    \includegraphics[width=0.45\linewidth]{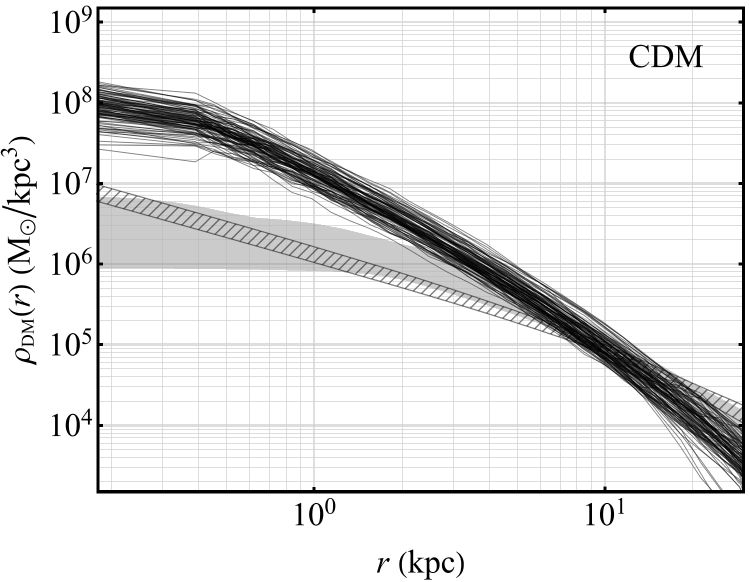}~~~~~~~~
    \includegraphics[width=0.45\linewidth]{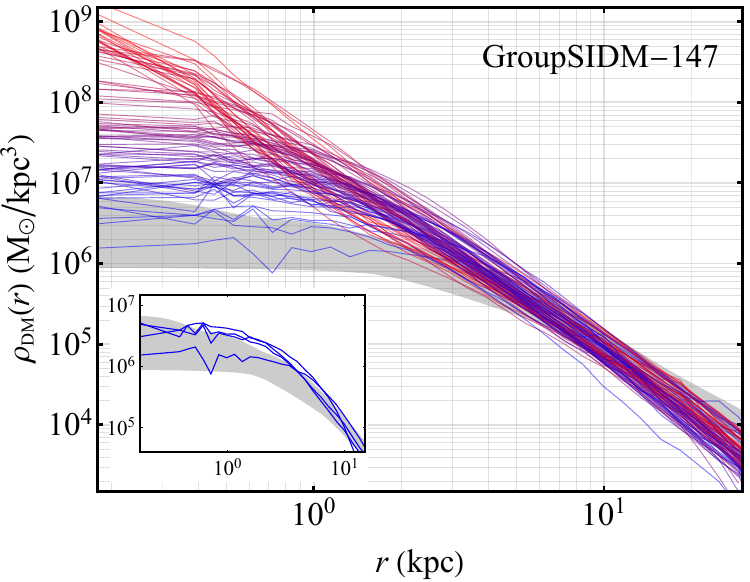}
    \caption{Density profiles of CDM and SIDM halos from the Concerto suite~\citep{Nadler:2023nrd}, in the left and right panels, respectively. The selected halos have masses in the range $(1.6\text{--}6.0)\times10^9\,\Ms$, encompassing the viable range of halo masses inferred for Cloud-9. The gray bands show the range of dark matter density profiles reconstructed from the $2\sigma$ parameter region of our MCMC analysis using the parametric SIDM halo model. In the left panel, the hatched band shows the $2\sigma$ range from the MCMC run with the NFW fit. The inset in the right panel highlights several SIDM halos compatible with Cloud-9.}
    \label{fig:simulations}
\end{figure*}

Previous studies based on the  CDM framework have found that Cloud-9-like systems should mostly reside beyond $300\textup{--}500\,{\rm kpc}$ from Milky Way-sized main halos~\citep{benitez-llambay_2017, Zheng:2025iyl, Guacimara:2026}. Thus, to find Cloud-9 analogs, we focus on simulated halos in isolation and select those with masses in the range $(1.6\textup{--}6.0)\times10^9 \, \Ms$, which brackets the viable range of halo masses from our MCMC analysis. In total, we identify $97$ halos for CDM and $93$ for SIDM. The distances of these halos from the center of the main halo (Halo004) are larger than $366\,{\rm kpc}$, and the majority lie between $2\textup{--}6\,{\rm Mpc}$. The density profiles for the selected CDM and SIDM halos are shown in the left and right panels in Fig.~\ref{fig:simulations}, respectively. For comparison, we reconstruct the $2\sigma$ range of the density profiles for Cloud-9 from the inferred parameters, denoted by the gray bands in Fig.~\ref{fig:simulations}.

From Fig.~\ref{fig:simulations}, we see that all CDM halos have significantly higher densities than those inferred for Cloud-9 within the inner region $r\lesssim 1\,{\rm kpc}$. In contrast, several SIDM halos have density profiles are compatible with Cloud-9 as highlighted in the inset of the right panel. These halos are still in the core-expansion phase and have a large density core.  For the GroupSIDM-147 model, the effective cross section is $\sigma/m\approx100\,{\rm cm^2/g}$~\citep{Yang:2022hkm,Yang:2022zkd} for the halo mass scale of Cloud-9, which is aligned with the favored value inferred from our MCMC analysis, see Fig.~\ref{fig:minicorner}. With such a large effective cross section, we expect that many of the SIDM halos, which have high concentrations, should be in a deeply core-collapse phase and have densities higher than their CDM counterparts, which is indeed what we see in the right panel of Fig.~\ref{fig:simulations}. 

Since Cloud-9 contains no detected stellar component, baryonic processes associated with galaxy formation are expected to have little or no impact on its dark matter halo. It therefore provides an exceptionally clean test of the underlying dark matter model. Although a cuspy NFW halo can reproduce the observed gas profile, doing so requires an anomalously low halo concentration, approximately $7\sigma$ below the cosmological median. By contrast, in SIDM models with large self-interacting cross sections, halos with only moderately low concentrations ($3\sigma$ below the median) naturally remain in the core-formation phase, producing large density cores that are consistent with the Cloud-9 observations. This picture is supported by both our MCMC analysis based on the parametric SIDM model and our cosmological N-body simulations.

It is interesting to note that the inferred properties of Cloud-9 resemble those of gas-rich ultra-diffuse galaxies (UDGs)~\citep{ManceraPina:2019zih,ManceraPina:2020ujo,ManceraPina:2024ybj}. Although these galaxies contain detectable stellar populations, their stellar masses are smaller than their gas masses, and the gas dominates the baryonic content. Their host halos are also inferred to have relatively low concentrations~\citep{Kong:2022oyk}. Gas-rich UDGs typically reside in halos with masses of a few times $10^{10}\,\Ms$, an order of magnitude larger than the halo masses expected for RELHICs such as Cloud-9. In this sense, gas-rich UDGs may represent a galaxy population adjacent to RELHICs along the sequence of low-mass, gas-dominated systems. Exploring the connection between RELHICs and gas-rich UDGs may therefore provide further insight into the formation and evolution of both populations.

As shown in the right panel of Fig.~\ref{fig:simulations}, many high-concentration halos within the relevant mass range are in the core-collapse phase and exhibit higher inner densities than their CDM counterparts. This raises the intriguing question of whether such core-collapsed halos can host RELHICs. As discussed in Sec.~\ref{sec:stable}, the gas distribution in compact halos with high central densities is likely to be unstable, potentially leading to gas collapse and star formation. Therefore, within the SIDM framework, halos with masses of a few times $10^9\,\Ms$ in the core-expansion phase can naturally host starless RELHICs like Cloud-9, whereas galaxies embedded in core-collapsed halos may evolve into stellar systems. We leave a detailed investigation of this possibility to future work.

\section{Conclusion}

Cloud-9 provides a clean testbed for probing the dark matter distribution. Assuming it is a RELHIC, the observed gas profile can be used to reconstruct the halo density profile through hydrostatic equilibrium, enabling constraints on the particle nature of dark matter. In this work, we carried out a comprehensive analysis in both the CDM and SIDM frameworks by fitting the observational data, requiring dynamical stability of the gas distribution, and reconstructing the posterior distributions of the halo parameters.

We identified a narrow degeneracy between the halo concentration and the stage of gravothermal evolution. Although halos spanning a wide range of evolutionary stages can reproduce the observed gas profile, incorporating the cosmological concentration--mass relation strongly favors SIDM halos in the core-forming phase. In particular, for $\sigma/m\gtrsim50\,{\rm cm^2/g}$, the preferred halos lie within $3\textup{--}4\sigma$ of the cosmological median concentration, whereas the CDM limit requires an exceptionally low-concentration halo, approximately $7\sigma$ below the median. Thus, while Cloud-9 alone cannot uniquely distinguish CDM from SIDM based on the gas profile, cosmological considerations strongly favor an SIDM interpretation with large self-interacting cross sections. We further searched our cosmological zoom-in simulation suite for Cloud-9 analogs and identified several SIDM candidates, but no CDM halo consistent with the inferred halo properties.

Our results demonstrate that RELHICs provide a powerful new probe of dark matter self-interactions that is complementary to traditional probes based on galaxy rotation curves, stellar kinematics, and strong gravitational lensing. As additional RELHIC candidates are discovered, population studies will enable increasingly stringent tests of SIDM and other dark matter models. On the theoretical side, hydrodynamical simulations of RELHIC formation in SIDM halos, particularly those undergoing gravothermal collapse, will provide a more complete understanding of the interplay between gas and dark matter and further sharpen the constraints on the particle nature of dark matter.

\section*{Code and Data Availability}
The code used for this calculation and analysis is publicly available at \href{https://github.com/morgan-ohana/Cloud9}{github.com/morgan-ohana/Cloud9}. In particular, the Rust implementation of the stretch-move MCMC algorithm developed for this project is available for general purpose use at \href{https://crates.io/crates/ensemble-mcmc}{crates.io/crates/ensemble-mcmc}. We also provide the Rust-native corner plot library developed for this analysis at \href{https://crates.io/crates/corner-plot}{crates.io/crates/corner-plot}. The Concerto dataset is available at \href{https://zenodo.org/records/14933624}{zenodo.org/records/14933624}.

\section*{Acknowledgments}
We thank Demao Kong for many helpful discussions. This work was supported by the U.S. Department of Energy under grant No. DE-SC0008541 and the John Templeton Foundation under grant \# 63599. The opinions expressed in this publication are those of the authors and do not necessarily reflect the views of the funding agencies.
\printcredits

\bibliographystyle{cas-model2-names}
\bibliography{ref}

\end{document}